\documentclass[twocolumn,aps,prl,superscriptaddress,showkeys,showpacs]{revtex4-1}

\usepackage{graphicx}
\usepackage{times}
\usepackage{setspace}

\begin{document}


\title{Relativity and the lead-acid battery}

\author{Rajeev Ahuja}
\email[]{rajeev.ahuja@fysik.uu.se }
\affiliation{Division of Materials Theory, Department of Physics and Astronomy, Uppsala University, Box 516, SE-751 20, Uppsala, Sweden}

\author{Andreas Blomqvist}
\affiliation{Division of Materials Theory, Department of Physics and Astronomy, Uppsala University, Box 516, SE-751 20, Uppsala, Sweden}

\author{Peter Larsson}
\affiliation{Division of Materials Theory, Department of Physics and Astronomy, Uppsala University, Box 516, SE-751 20, Uppsala, Sweden}

\author{Pekka Pyykk\"{o}}
\email[]{pekka.pyykko@helsinki.fi}
\affiliation{Department of Chemistry, University of Helsinki, Box 55 (A. I. Virtasen aukio 1), FI-00014 Helsinki, Finland}

\author{Patryk Zaleski-Ejgierd}
\email[]{patryk.ze@cornell.edu}
\affiliation{Department of Chemistry, University of Helsinki, Box 55 (A. I. Virtasen aukio 1), FI-00014 Helsinki, Finland}


\date{\today}

\begin{abstract}
The energies of the solid reactants in the lead-acid battery are
calculated ab initio using two different basis sets at
non-relativistic, scalar relativistic, and fully relativistic
levels, and using several exchange-correlation potentials. The average calculated standard
voltage is 2.13~V, compared with the experimental value of 2.11~V.
All calculations agree in that 1.7-1.8~V of this standard voltage
arise from relativistic effects, mainly from PbO$_2$ but also from
PbSO$_4$.
\end{abstract}



\maketitle


The lead battery is an essential part of cars, and has numerous
other applications. This well-known invention is now 150 years old
\cite{Plante:60, Kurzweil:10}.
About 75\% of the World lead production and a turnover of about 30
billion USD are due to these batteries. Although there are
electrochemical simulations starting from the given thermodynamical
data\cite{10,11}, we are not aware of any  $ab$ $initio$ ones for
the lead battery. This is in stark contrast to other rechargeable
batteries, such as the modern lithium-ion based systems, where they
abound. The problem is difficult enough to be a theoretical
challenge, and there is the additional fascination that, Pb being a
heavy-element, relativistic effects on its compounds could play an
important role, as qualitatively found a long time
ago\cite{Phillips:73,Desclaux:74b,1,2}. For metallic lead, see
\cite{Christensen:86,Liu:91,Verstraete:08}.

The discharge reaction of the lead-acid cell is
\begin{eqnarray}
 {\rm Pb}(s)+{\rm PbO{_2}}(s)+2{\rm H{_2}SO{_4}}(aq)\rightarrow \\ \nonumber
 2{\rm PbSO_4}(s)+2{\rm H_2O}(l),\hspace{0.75cm}\Delta E(1).
\end{eqnarray}

The electronic structures of both PbO\cite{4,5,6} and $\beta$-PbO$_2$\cite{7,8,9}
have been theoretically studied earlier. Especially, the metallic conductivity of
the $\beta$-PbO$_2$, making the large currents possible,
was shown to be an impurity, conduction-band effect, attributed to donor impurities at oxygen sites\cite{Anderson:65,Loucks:65,8,9}.
The alloying of the Pb electrode is also important in practice, but is
not discussed here, because the minute amounts of other elements do
not affect the EMF of the cell.

The construction of the lead-acid battery\cite{12} has a positive lead dioxide electrode,
a negative electrode of metallic lead, and a sulfuric acid electrolyte.
The discharge reaction between a Pb(IV) and a Pb(0) produces 2 Pb(II),
in form of solid PbSO$_4$. The experimental thermodynamics of the reaction are
well-known\cite{13}.


The three solids can be treated with existing solid-state theories, such as density functional theory (DFT), because the bonding mechanism in the investigated species is
dominated by covalent interactions where DFT is expected to provide
reliable results. Adequately simulating the liquid phase in multiple relativistic regimes is beyond current state of the art, however. We avoid this problem by introducing the known energy $\Delta$E(\ref{E2}) for the experimental reaction
\begin{equation}
 {\rm H_2O}(l)+{\rm SO_3}(g)\rightarrow~ {\rm H{_2}SO{_4}}(l)+\Delta E(\ref{E2}).\label{E2}
\end{equation}

\noindent We can use this empirical relationship because only light elements and only S(VI)  occur in eq.~(\ref{E2}), whose contribution to relativistic effects are small. Combining the equations (1) and (\ref{E2}) gives
\begin{equation}
 {\rm Pb}(s)+{\rm PbO_2}(s)+2{\rm SO_3}(g)\rightarrow~2{\rm Pb{_2}SO{_4}}(s)+\Delta E(\ref{E3}).\label{E3}
\end{equation}

\noindent The voltages for the lead-acid battery reaction may then be calculated from the reaction energies
\begin{equation}
 \Delta E(1) = \Delta E(\ref{E3})-2\Delta E(\ref{E2})\label{E4}
\end{equation}

\noindent where we use calculated $\Delta$E(\ref{E3}) values and experimental
$\Delta$E(\ref{E2}) values. Concentrated sulfuric acid is used in reactions (1) and
(\ref{E2}). The cell voltage at typical 5.5M (in H$_2$SO$_4$$\cdot$10H$_2$O) is calculated
using the values tabulated by Duisman\cite{13}.



Prediction of formation energies in a quantitative manner from ab initio calculation requires, in additional to having an accurate underlying theory, also absolute convergence of all technical parameters and a sufficiently general basis set. It is therefore fruitful to approach the problem with several independent methods, and see if they converge on the result. In our case, we used a linear combination of local orbitals (LCAO), with and without a frozen core approximation, using the BAND program\cite{17}; and a full-potential local-orbital minimum-basis approach\cite{18} with the FPLO program (version 7.00-27).

The calculations are performed with crystal structures from experimental room-temperature measurements (for structures see the Supplementary Material), allowing no ionic relaxations, thus
capturing the dynamic electronic effects of relativity (meaning Dirac
vs. Schr\"{o}dinger). The alternative, more laborious choice would
have been to also consider the relativistic structural and
vibrational changes, including zero-point energies (ZPE). The
thermal effects on the reaction energies are fairly small and are
simply neglected. If a battery freezes at low temperatures, it is
due to the kinetics, not due to the $\Delta$G. As stated, we assume
for simplicity pure, unalloyed Pb and a pure $\beta$-PbO$_2$ phase
in the electrode materials. The measured cell voltages for $\alpha$
and $\beta$-PbO$_2$ differ by ca. 0.01 V\cite{15}.

In the LCAO approach, the one-electron basis sets representing the optimized
electron density consist of linear combination of Herman-Skillman
numerical atomic orbitals (NO) and Slater-type orbitals (STO). We
apply a quadruple-zeta basis sets augmented with four polarization
functions (QZ4P), available in the BAND basis-set repository.
Frozen-core approximation is consistently applied to reduce the size
of the variational basis set. The use of frozen core, as implemented
in BAND\cite{17}, is preferable over pseudopotential techniques
because it essentially allows for all-electron calculations. The
frozen core orbitals are taken from high-accuracy calculations with
extensive STO basis sets. For oxygen and sulphur, we use all-electron
basis sets. For lead we include up to 4f orbitals in the core. The
slight change in the deep-core orbitals due to formation of chemical
bonds was confirmed to be negligible.

The band nature of the solid-state species was inferred by studying
the density of states (DOS). The orbital character of the total DOS
was determined with respect to contributions from individual atoms.
The band structure along a series of lines of high symmetry was also
analyzed. The overall band structures change only slightly
with the method: the ordering, the orbital character and the
dispersion remain comparable within LDA and GGA approximations.

In the case of BAND calculations, we used experimentally determined
structures both for solid- and gas-phase species. The lattice
constants and atomic positions were kept fixed in all calculations, motived by our interest in purely the electronic rearrangement effects due to relativity. The relativistic effects are investigated by means of the zeroth-order regular approximation (ZORA). We
consider three cases: non-relativistic (NR), scalar relativistic
(SR) and the fully relativistic (FR) with first-order spin-orbit
effects taken into account. To ensure high accuracy results, the
convergence of the calculations was checked with respect to all
crucial numerical parameters including the number of k-points, basis
set quality and size of the frozen core. The formation energies were
calculated with respect to spherically symmetric spin-restricted
atoms and converged within 1 kJ/mol or less. To sample the first
Brillouin zone, and evaluate the k-space integrals, we used a
quadratic numerical integration scheme with 9$\times$9$\times$9 (Pb
and Sn), 5$\times$5$\times$5 (PbO, PbO$_2$, SnO, SnO$_2$) and
3$\times$3$\times$3 (PbSO$_4$, SnSO$_4$) meshes. The SO$_3$ molecule
was placed in the middle of a large cubic unit cell (a=20 \AA) with a single k-point.

In the FPLO approach, we performed band-structure calculations using the
non-relativistic, the scalar-relativistic and the full
four-component relativistic versions of the full potential local
orbital minimum-basis band-structure method\cite{18} (FPLO version
7.00-27). The Local density approximation was used\cite{22} for the
exchange-correlation functional. Here, the optimal volumes of the crystal
structures were determined by calculating the experimentally
determined structures at different volumes and fitting a 4th-order
polynomial equation of state to the data points. To sample the first
Brillouin zone and evaluate the k-space integrals we use a
linear numerical integration scheme with 24$\times$24$\times$24 (Pb
and Sn), 12$\times$12$\times$12 (PbO, PbO$_2$, SnO, SnO$_2$) and
6$\times$6$\times$6 (PbSO$_4$, SnSO$_4$) k-point meshes. The SO$_3$ molecule
was calculated without periodic boundary conditions.

\begin{table}[t]
 \caption{Comparison of the experimental and calculated results for the EMF [V] of the lead-battery reaction (1).\label{T1}}
 \begin{tabular}{llrclcccr}
\hline\hline
\multicolumn{2}{l}{Method} & \multicolumn{3}{c}{Level of relativity}     &  $\Delta$     & $\Delta$      \\
\cline{3-5}
        &                  & NR           & SR      & FR                 & FR-NR & SR-NR \\
\hline
BAND    & VWN              &+0.55         &+2.52    &+2.27               &+1.72 &+1.97   \\
        & PBEsol-D         &+0.21         &+2.25    &+2.02               &+1.81 &+2.04   \\
\\
FPLO    & PW92             &+0.41         &+2.20    &+2.10               &+1.69 &+1.80   \\
        & PW92b            &+0.39         &+2.21    &+2.11               &+1.72 &+1.82   \\
\\
        & Av.              &+0.39         &+2.30    &+2.13               &+1.74 &+1.91   \\
        & Exp.$^a$         & & &+2.107                          &       &       \\
\hline\hline
 \end{tabular}
 \\$^a$ Ref.~\cite{15}
 \end{table}

\begin{figure}[b]
\includegraphics[width=86mm]{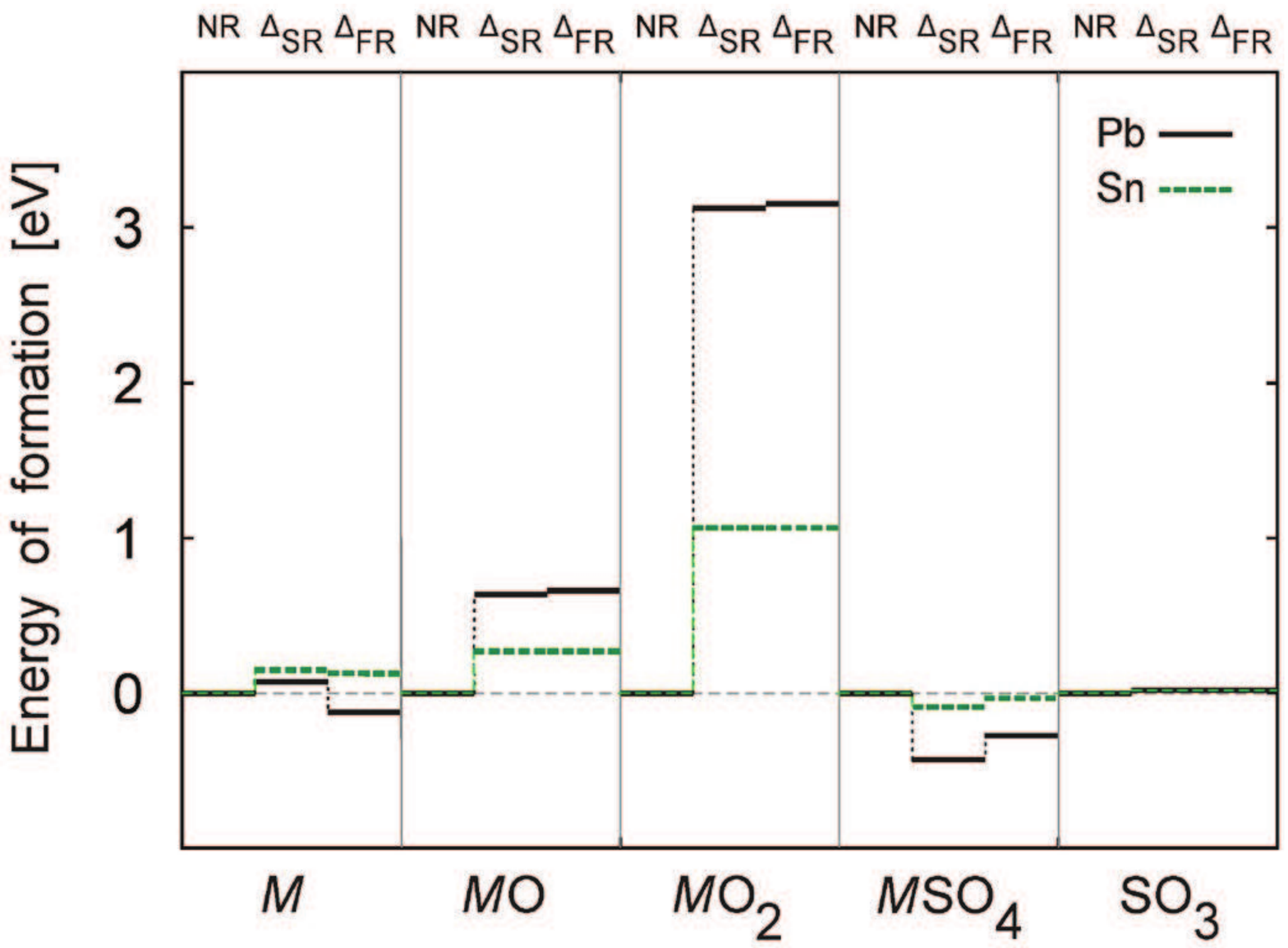}
\caption{The relativistic shifts in energies of formation, $E_f$
(per formula unit), calculated at DFT/PBEsol-D level.
The non-relativistic energy, NR = E$_f$(NR), is chosen as reference;
$\Delta_{\rm SR}$ = $E_f$(SR) - $E_f$(NR), and
$\Delta_{\rm FR}$ = $E_f$(FR) - $E_f$(NR).
Note the smallness of the effect on SO$_3$(g).}{\label{Pbf1}}
\end{figure}

The resulting reaction energies for the lead battery reaction(1), calculated at the FR (fully relativistic), SR (scalar relativistic = without spin-orbit coupling), and NR (non-relativistic)
levels are compared to experiment in Table \ref{T1}. We manage to reproduce the absolute voltage of the lead battery reaction within about 0.2V. Taking the four calculations in Table \ref{T1}
at face value, our calculated absolute voltage for reaction
(1), in H$_2$SO$_4$$\cdot$10H$_2$O, is +2.13 V while
its  relativistic part is +1.74 V. The relativistic increase of
the oxidative power of $\beta$-PbO$_2$(s) is indeed the largest contribution.
The PbSO$_4$(s) contribution to EMF follows, and has the same sign.
The third largest contribution comes from the spin-orbit coupling
effects in metallic Pb. At 'PBEsol-D' level of theory these three
contributions are +1.58, +0.27 and -0.06~V, respectively,
and were seen to be essentially method-independent. For details on individual
contributions, see Fig.~\ref{Pbf1}.

\begin{figure}[t]
\includegraphics[width=79mm]{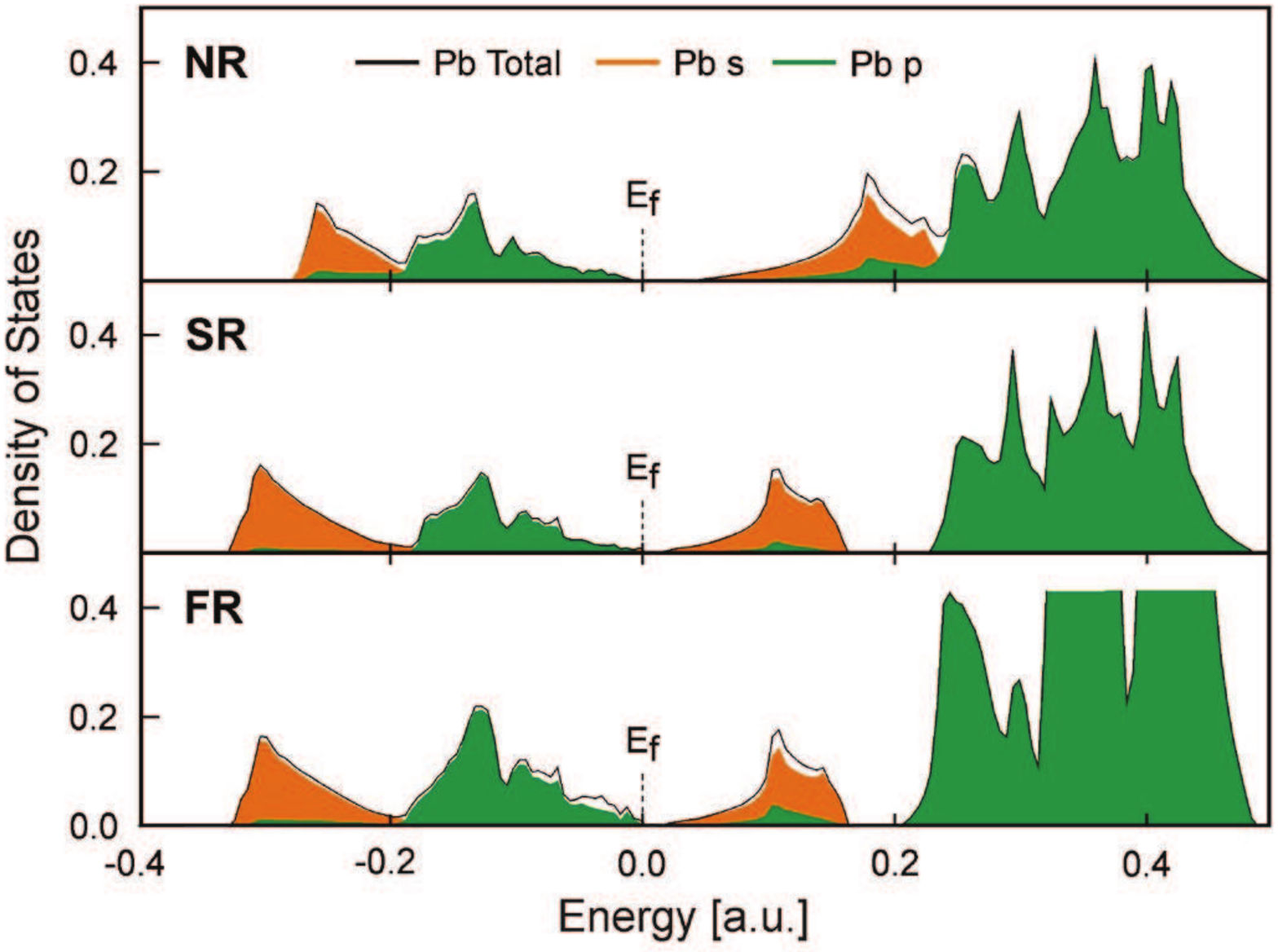}\vspace{4.5mm}
\caption{Total and partial DOS (states/eV/cell) of Pb in
$\beta$-PbO$_2$ calculated at DFT/VWN level (the Fermi energy,
$E_F$, is set to 0). Note the pronounced relativistic shift
in both occupied and unoccupied Pb 6s-states. (NR - non-relativistic,
SR - scalar-relativistic, FR - fully-relativistic).
Orange: Pb 6s, green: Pb 6p.\vspace{5mm}}{\label{Pbf2}}
\end{figure}

Comparing Pb with its lighter congener Sn,
it has been noticed before that no corresponding 'tin battery' exists\cite{16}.
This was attributed to the lower oxidative power of tin dioxide, as compared to
lead dioxide. We find support for this conclusion in our calculations, since a
tin battery would roughly correspond to a lead battery without relativistic effects
(see below). Indeed we found the largest relativistic shifts in the lead dioxide.
In Fig. \ref{Pbf2} we show the FR, SR, and NR densities of states (DOS) of Pb in
$\beta$-PbO$_2$(s). We see that the Pb 6s character is evenly distributed between
the filled and empty states\cite{8}, making this shell "half-oxidized". Furthermore,
the size of the band gap decreases with increasing level of relativity (NR, SR, FR).
Note also the relative shifts and changes in the Pb 6s population at different levels
of relativity. The Pb 6s states are significantly stabilized by inclusion of
relativity also in $\beta$-PbO$_2$.

We derived above the cell voltage of reaction (1)
by using a combination of experimental and theoretical  results. In order to validate the solid-state DFT  part - something which cannot be taken for granted - we also considered the
simplified 'toy model' reactions, taking place entirely in the solid-state,
\begin{equation}
 {\rm M}(s)+{\rm MO_2}(s)\rightarrow~2 {\rm MO}(s), \label{E5}
\end{equation}

\begin{figure}[t]
\includegraphics[width=87mm]{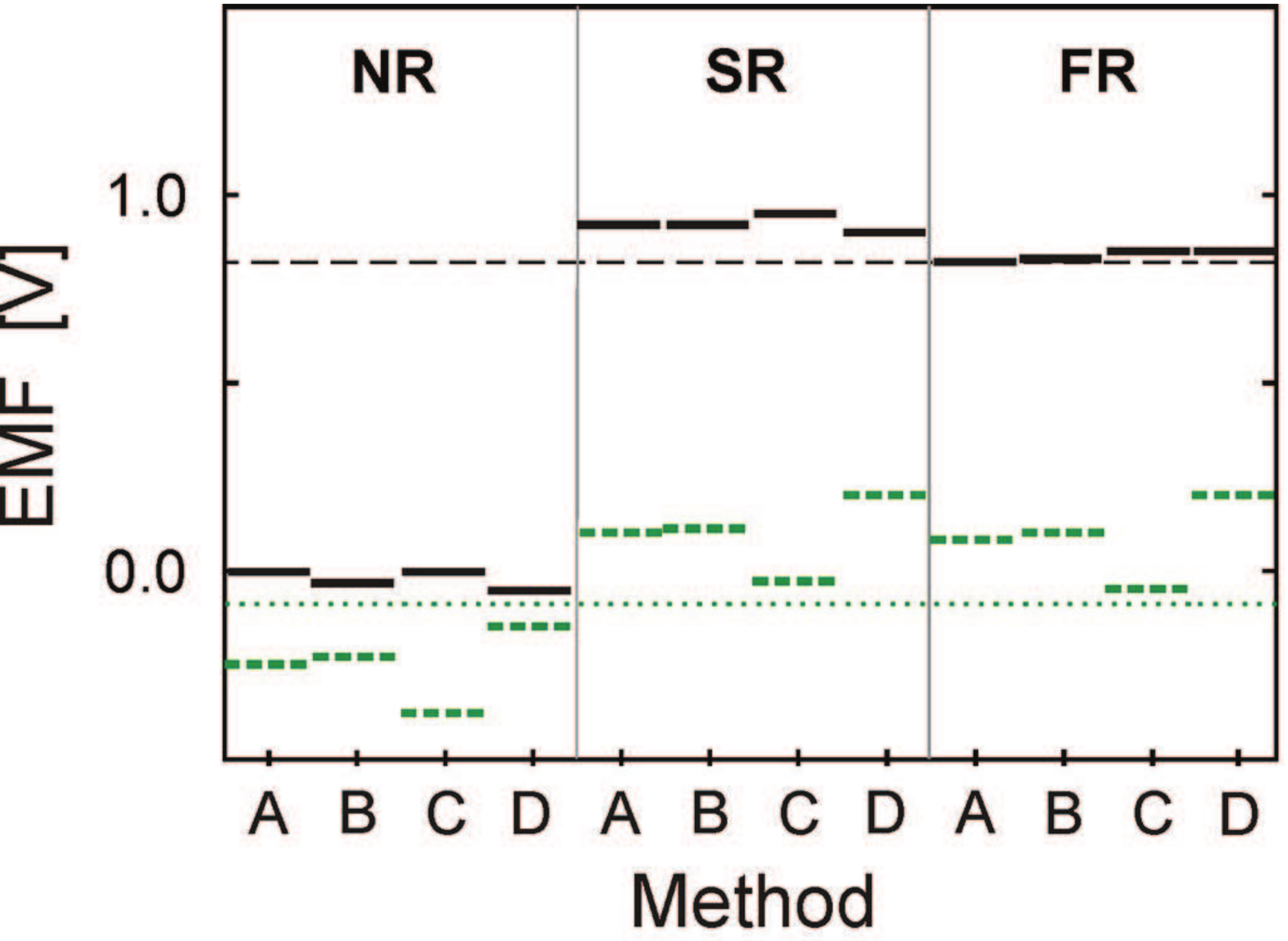}
\caption{Effect of relativity on the EMF of model-reaction,
$M$(s) + $M$O$_2$(s) $\rightarrow$ 2$M$O(s),
calculated with four different methods (A-D).
Experimental values for $M$=Pb (+0.82~V) and $M$=Sn (-0.085~V)
are indicated with the horizontal dashed and dotted lines,
respectively. A) BAND (VWN), B) BAND (PBEsol), C) BAND (PBEsol-D),
and D) FPLO (PW92). The level C has dispersion corrections.}{\label{Pbf3}}
\end{figure}

\noindent with M being Pb or Sn.  The reaction energy of this model
can be calculated completely by DFT, and should be close to experiment.
We find that this is the case, with the voltages for Pb being within 0.05~V
of the experimental value for the two different DFT implementations and several different exchange-correlation functionals, see Fig.~\ref{Pbf3}.
This suggests that our theoretical calculations accurately describe reality.
In Fig. \ref{Pbf1}, we show the calculated energies of formation, $E_f$, of
individual species at various levels of relativity. The relativistic shift
is most pronounced for solid $\beta$-PbO$_2$. For solid PbO it is
approximately five times smaller and has the same sign. In the
case of the tin 'toy-model', the largest relativistic contribution
arises from SnO$_2$, followed by SnO and Sn, for which the relativistic shifts
are only slightly different.

\begin{figure}[t]
\vspace{1.1cm}
\includegraphics[width=75mm]{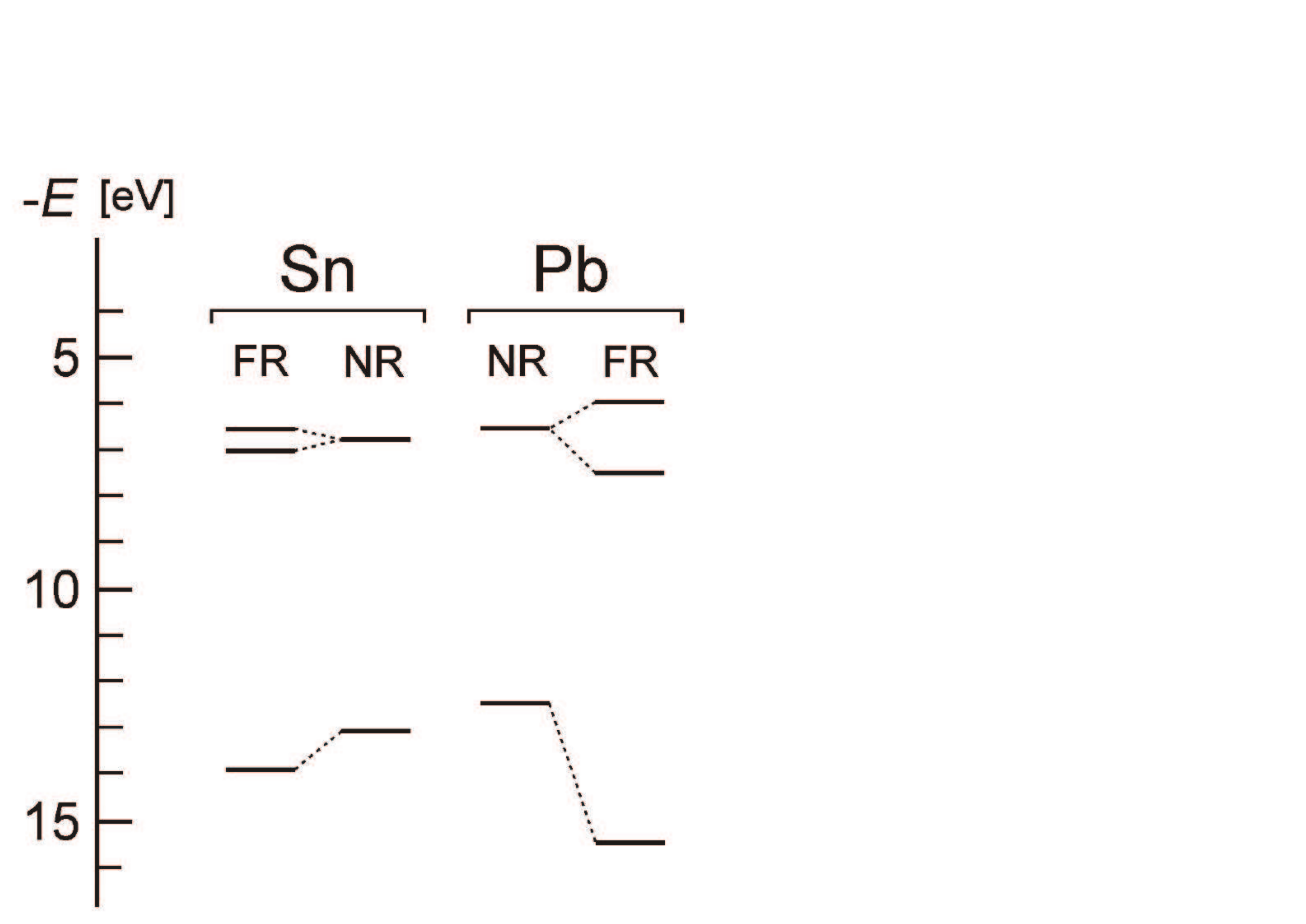}
\caption{Relativistic (FR) and non-relativistic (NR) Hartree-Fock orbital
energies for the tin and lead atoms in their $n$s$^2n$p$^2$  ground state.
Note the relativistic stabilization of the $n$s level (lowest in the figure),
leading to higher oxidative power for Pb(IV) than for Sn(IV).
Data from Desclaux\cite{Desclaux:73}.}{\label{Pbf4}}
\end{figure}

Returning to a qualitative discussion, we mean here by relativistic effects anything that depends on the speed of light or, more technically, the results of the Dirac versus Schr\"{o}dinger one-electron equation. The main effects are the stabilization of all $n$s and $n$p shells,
the destabilization of the $n$d and $n$f shells and the spin-orbit (SO)
splitting of the p,d,f, $\ldots$ shells. For example, the
 fully relativistic (FR) and non-relativistic (NR) atomic orbital energies
of tin and lead are shown in Fig.~\ref{Pbf4}. As noticed before\cite{1,2}, the non-relativistic values are similar for Sn and Pb while the relativistic ones are not. Qualitatively, the tendency
of Pb to be predominantly divalent can be related to the increased binding energy
of the Pb 6s shell.  Relativistic effects have also been suggested as responsible for the changing the structure of metallic lead from diamond to fcc\cite{Christensen:86,Hermann2010}, as well being necessary to determine phase transitions between fcc, hcp, and bcc structures\cite{Liu:91}.

Moreover, it has been shown that the valence-shell relativistic
effects of the Periodic Table scale roughly as $Z^2$, $Z$ being the full
nuclear charge\cite{2}. For tin and lead the ratio is
\begin{equation}
  \left [{{Z({\rm Sn})}\over{Z({\rm Pb})}}\right ]^2 = \left [{{50}\over{82}}\right ]^2 = 0.372, \label{E6}
\end{equation}

\noindent while the fully relativistic voltage changes, $\Delta E$,
calculated for the 'toy models'(\ref{E5}) yield
\begin{equation}
  \left [{{\Delta E({\rm Sn})}\over{\Delta E({\rm Pb})}}\right ]^2 = \left [{{0.34V}\over{0.86V}}\right ]^2 = 0.395. \label{E6b}
\end{equation}

\noindent Indeed, the EMF calculated for the non-relativistic lead 'toy-model' is similar to the analogous tin model reaction, treated at fully-relativistic level. It also explains why the analogous reaction (1) for tin is unknown.

Concluding, the lead-acid battery belongs to those familiar phenomena,
whose characteristic features are due to the relativistic dynamics of
fast electrons  when they move near a heavy nucleus. In this case the main actors are the 6s valence electrons of lead, in the substances involved.
This insight may not help one to improve the lead battery,
but it could be useful in exploring alternatives. Finally, we note that cars start due to relativity.

\begin{acknowledgments}
PP and PZE belong to the Finnish Center of Excellence in
Computational Molecular Science (CMS) and used computer resources at
CSC, Espoo, Finland. PZE acknowledges support by Magnus Ehrnrooths
Stiftelse and Finnish Cultural Foundation. RA, AB, and PL are
grateful for time allocated on Swedish National Supercomputing
facilities and for financial support from the Swedish Research
Council (VR) and Formas Research Council.
\end{acknowledgments}

\bibliography{rdataI_III,rdataIV,leadbattery}

\end{document}